# Density, speed of sound, refractive index and relative permittivity of methanol, propan-1-ol or pentan-1-ol + benzylamine liquid mixtures. Application of the Kirkwood-Fröhlich model

Fernando Hevia*[,a], Víctor Alonso[b], Ana Cobos[a], Juan Antonio González[a], Luis Felipe Sanz[a], Isaías García de la Fuente[a]

[a] G.E.T.E.F., Departamento de Física Aplicada. Facultad de Ciencias. Universidad de Valladolid. Paseo de Belén, 7, 47011 Valladolid, Spain.

[b] Departamento de Física Aplicada. EIFAB. Campus Duques de Soria. Universidad de Valladolid. C/ Universidad s.n. 42004 Soria, Spain.

*e-mail: luisfernando.hevia@uva.es

# Abstract

Densities ($\rho$), speeds of sound ($c$), relative permittivities at 1 MHz ($\varepsilon_r$) and refractive indices at the sodium D-line ($n_D$) at $T$ = (293.15 K to 303.15 K) and $p$ = 0.1 MPa are reported for binary liquid mixtures alkan-1-ol + benzylamine. Methanol, propan-1-ol and pentan-1-ol are the alkan-1-ols studied in this work. The values of the excess molar volume ($V_m^E$), excess isentropic compressibility ($\kappa_S^E$), excess speed of sound ($c^E$), excess refractive index ($n_D^E$), excess relative permittivity ($\varepsilon_r^E$) and its temperature derivative ($\partial\varepsilon_r^E / \partial T)_p$ are calculated, and they are adjusted to Redlich-Kister polynomials. The $V_m^E$ values are negative, indicating a predominance of the solvation between unlike molecules and structural effects. $\varepsilon_r^E$ values indicate a positive contribution from the creation of (alkan-1-ol)-benzylamine interactions, and the positive value for the methanol mixture emphasises the importance of solvation. Calculations on excess molar refractions point out to weaker dispersive interactions than in the ideal mixture, which may be explained by the mentioned solvation effects. The Kirkwood-Fröhlich model has been applied to the mixtures, and the Kirkwood correlation factors suggest an important relative weight, especially in the methanol system, of linear-like molecules in the solutions, which is in accordance with the positive contribution of the formed multimers to $\varepsilon_r^E$ due to their good effective response to the electric field.

# 1. Introduction

We are carrying out a systematic thermophysical characterization of binary mixtures containing alkan-1-ols and amines [1-16] with the aim to understand and model the interactions and structure of these liquid systems. In a series of previous works, we have provided volumetric, dielectric and refractive properties of systems containing cyclohexanamine [13,14], hexan-1-amine [10], *N*-propylpropan-1-amine [11], *N,N*-diethylethanamine [12] and aniline [15]. More recently, we have reported excess molar enthalpies of cyclohexanamine systems [16]. Now, we turn our attention to alkan-1-ol + benzylamine liquid mixtures. Among the numerous industrial uses of benzylamine, it is receiving attention in the field of carbon capture, utilization and storage, which is well-known to be a common research topic nowadays. Amine solutions seem to be very suitable for this type of application, and aqueous solutions of benzylamine could be candidates in this field. Recently, heats of absorption of $CO_2$ and heat capacities have been experimentally determined for this type of solutions [17]. From the theoretical perspective, it is interesting to investigate the effect of replacing aniline by benzylamine in systems with a given alkan-1-ol. While aniline and benzylamine differ only in a $-CH_2-$ group between the amine group and the aromatic ring that is present in the latter, their macroscopic properties are very different. In pure aniline there exist very strong dipolar interactions, while in pure benzylamine such interactions are weaker. This can be clearly shown by comparing the upper critical solution temperatures (UCST) of mixtures with alkanes. For instance, UCST(aniline + heptane) = 343.11 K [18], and UCST(benzylamine + decane) = 280.09 K [19].

In this article, we report experimental densities ($\rho$), speeds of sound (*c*), relative permittivities at 1 MHz ($\varepsilon_r$) and refractive indices at the sodium D-line ($n_D$) for methanol, propan-1-ol and pentan-1-ol + benzylamine liquid mixtures at pressure *p* = 0.1 MPa and in the temperature range *T* = (293.15 to 303.15) K, and we calculate the Kirkwood correlation factors in the framework of the one-fluid model version of the Kirkwood-Fröhlich theory [20,21] in order to gain insight into the correlations between the molecular dipoles in the studied solutions.

# 2. Experimental

## 2.1. Materials

Pure liquids were used without further purification. Information about their source and purity is displayed in Table 1. Table 2 lists the measured properties of the pure compounds, whose agreement with literature values is analysed below, and their dipole moments.

## 2.2.  Apparatus and procedure

The concentration of the liquid mixtures was calculated from mass measurements. The masses were determined by weighing using a Sartorius MSU125p analytical balance and correcting for buoyancy effects, with a standard uncertainty of $5 \cdot 10^{-5}$ g. The corresponding standard uncertainty of the mole fraction is 0.0005. Some cautions were taken: i) To minimize interaction with air, pure liquids were stored with 4 Å molecular sieves, except for methanol, which interacted with the sieves; ii) the density of the pure compounds was measured over time to check their stability, staying constant within the uncertainty of the measurements; iii) the measurement cells were appropriately filled and closed to avoid partial evaporation.

Pt-100 resistances, calibrated using the triple point of water and the melting point of Ga, were used to measure the temperature of the samples, with a standard uncertainty of 0.01 K for $\rho$ and $c$, and 0.02 K for $\varepsilon_r$ and $n_D$.

A densimeter and sound analyser from Anton Paar, model DSA 5000, was employed to determine densities (by the vibrant tube method) and speeds of sound (using ultrasonic pulses at 3 MHz centre frequency) of liquid samples. with a temperature stability of 0.001 K. More details about the technique and the calibration and test of the apparatus can be found elsewhere [15,22,23].

The refractive property $n_D$ was determined using an automatic refractometer Bellingham + Stanley RFM970. The temperature is regulated with Peltier modules with a stability of 0.02 K. The apparatus was calibrated with 2,2,4-trimethylpentane and toluene at $T$ = (293.15 to 303.15) K, with reference values recommended by Marsh [24]. For 2,2,4-trimethylpentane, the values are 1.39145 (293.15 K), 1.38898 (298.15ºC) and 1.38650 (303.15 K); for toluene, 1.49693 (293.15 K), 1.49413 (298.15 K) and 1.49126 (303.15 K).

The experimental device to determine $\varepsilon_r$ and its calibration has been described in detail elsewhere [25]. The test of the technique and the uncertainty assessment were aided by the comparison with literature values of many pure liquids at $T$ = (288.15 to 333.15) K [12]. The procedure involves precise impedance measurements of a parallel-plate capacitor (Agilent 16452A, $\approx 4.8$ cm$^3$) with an impedance analyser (Agilent 4294A). The temperature of the cell is regulated by a thermostatic bath (LAUDA RE304) with a stability of 0.02 K.

## 2.3.  Uncertainty of the measurements

The uncertainty assessed from measurements of different samples of the same liquid (but always prepared from the same source liquids), carried out by us, using the measurement procedures and instruments described above, under the same conditions, in the same location and repeated over a short period of time will be called *repeatability*. This term will be therefore referred to our measurements only. It will also include the uncertainty associated to calibration constants and/or reference values where appropriate, but it will not consider the purity of the source liquids. For the measurements reported in this work, the expanded uncertainties (with a coverage factor of 2, approximately 95% confidence level) associated with the repeatability were found to be: 0.0001 for the mole fraction, 0.00005 g·cm$^{-3}$ for $\rho$, 0.8 m·s$^{-1}$ for $c$, 0.00008 for $n_D$, and 0.001$\varepsilon_r$ (0.1 % relative expanded uncertainty) for $\varepsilon_r$. These values were used to define the number of digits reported for each of the properties.

However, the total uncertainty associated with these quantities should ideally include other factors, such of the purity of the compounds, which is very hard to quantify. For the mole fraction of the mixtures, the total expanded uncertainty has been roughly estimated as 0.0010. For the other properties mentioned, this total uncertainty has been quantified by the *reproducibility*, which we will take as the uncertainty evaluated from measurements of the same liquid carried out by different authors –and therefore, different source liquids, principles, methods, instruments, locations, conditions and/or periods of time–. For the case of alkan-1-ols, the agreement of the measured properties by different authors is generally better than for benzylamine. For the latter case, we include in Table 2 more literature values in order to analyse the differences and assess a reasonable estimate of the total uncertainty. Some of them have been considered as outliers (the $\rho$ value from reference [26], the $\rho$ and $c$ values from reference [27,28], and the $n_D$ values from reference [29]) due to their unusually large differences to the corresponding average values. For the relative permittivity of benzylamine, there are not enough data to perform a significant analysis (to the best of our knowledge, only one reference is available [85]), so they have been ignored in the uncertainty assessment for $\varepsilon_r$. With the above considerations, the expanded uncertainties (with a coverage factor of 2) associated with the reproducibility have been estimated as: 0.0006 g·cm$^{-3}$ for $\rho$, 1.3 m·s$^{-1}$ for $c$, 0.0003 for $n_D$, and 0.01$\varepsilon_r$ (1 % relative expanded uncertainty) for $\varepsilon_r$.

The uncertainties of the excess functions (defined below) are given in Tables 3-7. They are estimated from repeatability considerations and/or comparison with reference values of test systems, such as cyclohexane + benzene for the excess molar volume, excess isentropic compressibility and excess speed of sound [30-32]. The estimation of the uncertainty of these functions using the reproducibility would give some unrealistically large uncertainties.

## 3. Equations

If dispersion and absorption of the acoustic wave are negligible, the Newton-Laplace equation allows to determine the isentropic compressibility ($\kappa_S$) from $\rho$ and $c$ experimental measurements:

$$\kappa_S = \frac{1}{\rho c^2} \tag{1}$$

The isothermal compressibility ($\kappa_T$) can be calculated through general thermodynamic relations involving $\kappa_S$, the molar isobaric heat capacity ($C_{pm}$), the molar volume ($V_m$) and the isobaric thermal expansion coefficient ($\alpha_p$):

$$\kappa_T = \kappa_S + \frac{T V_m (\alpha_p)^2}{C_{pm}} \tag{2}$$

The values $F^{id}$ of the thermodynamic properties of an ideal mixture at the same temperature and pressure as the real mixture are computed from the well-established formulae from Benson and Kiyohara [33-35]:

$$F^{id} = x_1 F_1^* + x_2 F_2^* \qquad (F = V_m, C_{pm}) \tag{3}$$

$$F^{id} = \phi_1 F_1^* + \phi_2 F_2^* \qquad (F = \alpha_p, \kappa_T) \tag{4}$$

where $F_i^*$ is the value of the property $F$ of pure component $i$, $x_i$ represents the mole fraction of component $i$ and $\phi_i = x_i V_{mi}^* / V_m^{id}$ is the volume fraction of component $i$. Ideal values of $\kappa_S$ and $c$ are calculated using the equations [33]:

$$\kappa_S^{id} = \kappa_T^{id} - \frac{T V_m^{id} (\alpha_p^{id})^2}{C_{pm}^{id}} \tag{5}$$

$$c^{id} = \left( \frac{1}{\rho^{id} \kappa_S^{id}} \right)^{1/2} \tag{6}$$

being $\rho^{id} = (x_1 M_1 + x_2 M_2) / V_m^{id}$ ($M_i$, molar mass of the $i$ component).

The ideal dielectric and refractive properties are calculated from [36,37]:

$$\varepsilon_r^{id} = \phi_1 \varepsilon_{r1}^* + \phi_2 \varepsilon_{r2}^* \tag{7}$$

$$n_D^{id} = \left[ \phi_1 (n_{D1}^*)^2 + \phi_2 (n_{D2}^*)^2 \right]^{1/2} \tag{8}$$

$$\left[ \left( \frac{\partial \varepsilon_r}{\partial T} \right)_p \right]^{id} = \left( \frac{\partial \varepsilon_r^{id}}{\partial T} \right)_p \tag{9}$$

Lastly, the excess properties are defined by:

$$F^E = F - F^{id} \tag{10}$$

## 4. Results

The $\alpha_p = -(1/\rho)(\partial\rho/\partial T)_p$ values of the pure compounds at $T = 298.15$ K and $p = 0.1$ MPa were obtained from fittings of experimental $\rho$ values to straight lines as functions of $T$ in the range (293.15 to 303.15) K. An analogous procedure was used for the temperature derivative $(\partial\varepsilon_r^E/\partial T)_p = [(\partial\varepsilon_r/\partial T)_p]^E = (\partial\varepsilon_r/\partial T)_p - (\partial\varepsilon_r^{id}/\partial T)_p$.

The experimental values of the determined quantities at $p = 0.1$ MPa for alkan-1-ol (1) + benzylamine (2) liquid mixtures can be found in Tables 3-7 as functions of $x_1$ (mole fraction of the alkan-1-ol). Values of $\rho$, $c$, and $V_m^E$ at $T = (293.15$ to 303.15) K are collected in Table 3; $\kappa_S^E$ and $c^E$ values at $T = 298.15$ K are written in Table 4; $\phi_1$, $\varepsilon_r$ and $\varepsilon_r^E$ values at $T = (293.15$ to 303.15) K are found in Table 5; whereas $\phi_1$, $n_D$ and $n_D^E$ values at the same conditions are written in Table 6. $(\partial\varepsilon_r^E/\partial T)_p$ results at $T = 298.15$ K are collected in Table 7.

The $F^E$ data were fitted by unweighted linear least-squares regressions to Redlich-Kister polynomials [38]:

$$F^E = x_1(1-x_1)\sum_{i=0}^{k-1} A_i(2x_1-1)^i \tag{11}$$

The number, $k$, of appropriate coefficients for each system, property and temperature has been determined by the application of an F-test of additional term [39] at a 99.5% confidence level. Table 8 includes the parameters $A_i$ obtained, and the standard deviations, $\sigma(F^E)$, defined by:

$$\sigma(F^E) = \left[\frac{1}{N-k}\sum_{j=1}^{N}(F_{cal,j}^E - F_{exp,j}^E)^2\right]^{1/2} \tag{12}$$

where $j$ indexes the $N$ experimental data $F_{exp,j}^E$, and $F_{cal,j}^E$ is the corresponding value of the excess property $F^E$ calculated from equation (11).

The values of the excess properties at $T = 298.15$ K and $p = 0.1$ MPa ($V_m^E$, $\kappa_S^E$ and $c^E$ versus $x_1$; and $\varepsilon_r^E$, $(\partial\varepsilon_r^E/\partial T)_p$ and $n_D^E$ versus $\phi_1$) are plotted in Figures 1-6 with the corresponding Redlich-Kister fittings.

## 5. Discussion

In this section, if nothing else is specified, the values of the thermophysical properties are given at $T = 298.15$ K and $x_1 = 0.5$, except for dielectric properties, which will be given at $\phi_1 =$ 0.5. As before, we will denote by $n$ the number of carbon atoms of the alkan-1-ols and by $n$OH the alkan-1-ol with $n$ carbon atoms.

## 5.1. Excess molar volumes, isentropic compressibilities and speeds of sound

$V_m^E$/cm$^3$·mol$^{-1}$ values of $n$OH + benzylamine liquid mixtures (Figure 1) are negative, revealing a predominance of interactions between unlike molecules and structural effects: −1.543 (1OH), −1.035 (3OH), −0.765 (5OH). In the case of the mixture containing methanol, structural effects may be of free volume type, if one takes into account the large difference between the $\alpha_p$ values of the components of the system (Table 2). Values of $V_m^E$ increase with $n$, a trend that may be explained by: i) a higher contribution from the rupture of benzylamine-benzylamine interactions in the mixing process by longer alkan-1-ols due to their larger aliphatic surface, and ii) a more important contribution from $n$OH-benzylamine interactions created in mixtures with shorter alkan-1-ols, since the OH group is more sterically hindered in longer alkan-1-ols. A similar interpretation can be given to the negative values and parallel variation with $n$ which is observed for $\kappa_S^E$/TPa$^{-1}$ (Figure 2): −109.8 (1OH), −71.0 (3OH), −42.6 (5OH). These $\kappa_S^E$/TPa$^{-1}$ values are similar to those of $n$OH + aniline systems [15]: −109.7 (1OH), −72.5 (3OH), −37.3 (5OH). This is an interesting fact, as the $V_m^E$/cm$^3$·mol$^{-1}$ values for the aniline mixtures are higher than those of benzylamine solutions: −0.902 (1OH, [15]), −0.593 (3OH, [15]), −0.398 (4OH [40]), −0.241 (5OH, [15]). The relation between $V_m^E$ of benzylamine and aniline mixtures can be understood considering that in pure aniline there exist stronger dipolar interactions and that the breaking of such interactions contributes more positively to $V_m^E$. It is to be noted that large structural effects exist in alkan-1-ol + aniline systems, since from ethanol $H_m^E$ values are positive [41,42] and $V_m^E$ values are negative [15]. The positive $c^E$/m·s$^{-1}$ obtained for alkan-1-ol + benzylamine (Figure 3) are consistent with the above results on $\kappa_S^E$: 131.0 (1OH), 80.1 (3OH), 47.6 (5OH).

The excess molar isobaric expansion, $A_p = (\partial V^E/\partial T)_P$, is negative for the 1OH + benzylamine mixture (−2.7·10$^{-3}$ cm$^3$·mol$^{-1}$·K$^{-1}$) and very small and positive for the other two systems (8.5·10$^{-4}$ and 1.8·10$^{-4}$ cm$^3$·mol$^{-1}$·K$^{-1}$ for the 3OH and 5OH solutions, respectively). A similar trend is encountered for alkan-1-ol + aniline systems. Thus, $A_p$/cm$^3$·mol$^{-1}$·K$^{-1}$ = −2.8·10$^{-3}$ (1OH) [15,43]; 6.5·10$^{-3}$ (4OH) [40]; regarding these values, note that, as discussed in an earlier article [15], (i) they should be taken with caution because of the dispersion of the literature data, and (ii) reference [40] for the 4OH system seems to be the one that agrees best with our data. Negative $A_p$ values have been linked to the existence of structural effects [44], but have also been interpreted in terms of a decrease in the molar volume of complex formation, which overcompensates for the decrease in the extent of complex formation, and have been encountered, e.g., in amine + trichloromethane mixtures [45,46].

## 5.2. Excess relative permittivities

Negative contributions to the value of $\varepsilon_r^E$ appear when interactions between like molecules are broken, and either positive or negative ones when interactions between unlike molecules are created, depending on the effective response of the formed multimers to an electric field. Dominance of the rupture of interactions between like molecules occurs in $n$OH + alkane mixtures, e.g. heptane [9,47-49], or benzene [49,50]. For the latter, $\varepsilon_r^E$ values (at 293.2 K) are: $-1.28$ (1OH), $-2.15$ (3OH), $-2.78$ (4OH), $-2.83$ (6OH), $-2.23$ (8OH). The values of this property for $n$OH + benzylamine systems are: 0.524 (1OH), $-1.507$ (3OH), $-1.856$ (5OH). The benzylamine molecule can be obtained by adding a $-CH_2NH_2$ group to the phenyl ring, and the contribution from the rupture of interactions between molecules of the same species is expected to be more negative than in benzene solutions. The higher $\varepsilon_r^E$ results for the benzylamine mixtures and the positive value for the 1OH + benzylamine mixture indicate that the contribution from the created interactions upon mixing to $\varepsilon_r^E$ is positive. The variation of $\varepsilon_r^E$ with $n$ can be explained by the weaker and lower association of longer alkan-1-ols [10].

$n$OH + aniline systems show lower negative $\varepsilon_r^E$ values [15]: $-0.778$ (1OH), $-1.850$ (3OH), $-2.082$ (5OH). The $\varepsilon_r^E$ curves of these systems, as well as those of excess molar enthalpies and excess molar internal energies at constant volume, are known to be characterized by the dominance of the rupture of the strong dipolar interactions between aniline molecules and of the alcohol network [15]. As said above, such strong dipolar interactions are not so relevant in liquid benzylamine, and this affects strongly the observed $\varepsilon_r^E$ values. Therefore, one should expect a more important relative weight of the solvation effects between molecules of $n$OH and benzylamine. This is also suggested by the Kirkwood-Fröhlich model applied to our data (see below), because the molecular dipoles in the benzylamine mixtures are, according to this theory, more parallelly aligned in average. Moreover, the breaking of aniline-aniline interactions by alkan-1-ol molecules should give a more negative contribution to $\varepsilon_r^E$ than in the case of benzylamine-benzylamine interactions, which is also in accordance with the observed lower $\varepsilon_r^E$ values of aniline systems.

The $T$ derivative of the excess relative permittivity is negative for 1OH + benzylamine (excluding a short interval at low $\phi_1$), and positive for the 3OH and 5OH mixtures. The curves are practically identical to the ones for $n$OH + aniline systems, and an analogous interpretation can be given [15].

### 5.3. Molar refraction

The molar refraction, $R_m$, can be used to quantify the dispersive interactions [21,51], as it is proportional to the average electronic contribution to the polarizability from one molecule in a macroscopic sphere of liquid [20,21]:

$$R_m = \frac{n_D^2 - 1}{n_D^2 + 2} V_m = \frac{N_A \alpha_e}{3\varepsilon_0} \tag{16}$$

($N_A$ is Avogadro's constant and $\varepsilon_0$ the vacuum permittivity). For $n$OH + benzylamine mixtures at equimolar composition, $R_m$ /cm$^3$·mol$^{-1}$ = 21.1 (1OH), 25.9 (3OH), 30.6 (5OH). They increase with $n$ and are higher than the corresponding values in $n$OH + aniline systems [15]: 19.5 (1OH), 24.2 (3OH), 28.9 (5OH). Dispersive interactions are therefore slightly stronger in benzylamine systems and also increase with $n$.

The substitution of ideal values in equation (16) allows to calculate the ideal molar refraction, $R_m^{id}$, and the excess value $R_m^E = R_m - R_m^{id}$. For $n$OH + benzylamine mixtures at equimolar composition, $R_m^E$ /cm$^3$·mol$^{-1}$ = –0.44 (1OH), –0.29 (3OH), –0.22 (5OH). Therefore, dispersive interactions are weaker than if the molecular dipoles of different species did not interact. The increase of $R_m^E$ with $n$ is similar to what is observed in $n$OH + aniline mixtures [15]: –0.27 (1OH), –0.17 (3OH), –0.07 (5OH), and a similar interpretation can be given: it may be due to a more important solvation for small $n$ values. Moreover, $R_m^E$ ($n$OH + benzylamine) < $R_m^E$ ($n$OH + aniline). This fact may be interpreted as follows. The breaking of dipolar interactions of aniline on mixing might lead to an important positive contribution to $R_m$ due to a corresponding increase of the freedom of the valence electrons to move, which is expected to be much less relevant in benzylamine systems because dipolar interactions do not play such a decisive role in this amine.

### 5.4. Kirkwood-Fröhlich model

The Kirkwood-Fröhlich model assumes that the molecules are in a spherical cavity of an infinitely large dielectric and treats macroscopically the induced contribution to the total polarizability by means of the high-frequency permittivity, $\varepsilon_r^\infty$. Long-range dipolar interactions are included in a local field in the cavity, considering its surroundings as a continuous medium of permittivity $\varepsilon_r$. The Kirkwood correlation factor, $g_K$, gathers information about short-range interactions by quantifying the deviations of the relative orientation of a dipole with respect to its neighbors from randomness. For a one-fluid model of a mixture of polar liquids [52], $g_K$ can be determined using [20,21,52,53]:

$$g_{\mathrm{K}} = \frac{9k_{\mathrm{B}}TV_{\mathrm{m}}\varepsilon_0(\varepsilon_{\mathrm{r}} - \varepsilon_{\mathrm{r}}^{\infty})(2\varepsilon_{\mathrm{r}} + \varepsilon_{\mathrm{r}}^{\infty})}{N_{\mathrm{A}}\mu^2\varepsilon_{\mathrm{r}}(\varepsilon_{\mathrm{r}}^{\infty} + 2)^2} \qquad (17)$$

where $k_{\mathrm{B}}$ is Boltzmann's constant; $N_{\mathrm{A}}$, Avogadro's constant; $\varepsilon_0$, the vacuum permittivity; and $V_{\mathrm{m}}$, the molar volume of the liquid at the working temperature, $T$. For polar compounds, $\varepsilon_{\mathrm{r}}^{\infty}$ has been estimated using $\varepsilon_{\mathrm{r}}^{\infty} = 1.1n_{\mathrm{D}}^2$ [54]. $\mu$ represents the dipole moment of the solution, estimated from the equation [52]:

$$\mu^2 = x_1\mu_1^2 + x_2\mu_2^2 \qquad (18)$$

where $\mu_i$ stands for the dipole moment of component $i$ (= 1,2). Calculations have been performed using smoothed values of $V_{\mathrm{m}}^{\mathrm{E}}$, $n_{\mathrm{D}}^{\mathrm{E}}$ and $\varepsilon_{\mathrm{r}}^{\mathrm{E}}$ at $\Delta x_1 = 0.01$. The source and values of $\mu_i$ used for alkan-1-ol + benzylamine mixtures are collected in Table 2.

For the mixtures $n$OH + benzylamine under study, the $g_{\mathrm{K}}$ values decrease with $n$ (Figure 7): 2.22 (1OH), 1.90 (3OH). 1.63 (5OH), due to a weaker and lower solvation and self-association of the alkan-1-ol when it is longer. These $g_{\mathrm{K}}$ values are somewhat higher than the ones found for the corresponding aniline solutions [15]: 1.94 (1OH), 1.72 (3OH). 1.50 (5OH). As a sample for comparison, the curve for 1OH + aniline is represented in Figure 7. Therefore, as advanced above, the Kirkwood-Fröhlich model suggests a more parallel alignment of molecular dipoles in benzylamine mixtures than in aniline systems. This is an interesting fact, pointing out to a more important relative weight of linear-like multimers against cyclic-like ones in the mixtures containing benzylamine.

## 6. Conclusions

Densities, speeds of sound, relative permittivities at 1 MHz and refractive indices at the sodium D-line have been measured for methanol, propan-1-ol or pentan-1-ol + benzylamine liquid mixtures at 0.1 MPa and $T$ = (293.15 to 303.15) K. Negative values of $V_{\mathrm{m}}^{\mathrm{E}}$ are found, revealing a predominance of interactions between unlike molecules and structural effects. The analysis of $\varepsilon_{\mathrm{r}}^{\mathrm{E}}$ values reveals a positive contribution to this quantity from the creation of interactions between unlike molecules. The positive value for the methanol mixture underlines their importance in the systems under study. Alkan-1-ol + aniline systems show higher $V_{\mathrm{m}}^{\mathrm{E}}$ and lower $\varepsilon_{\mathrm{r}}^{\mathrm{E}}$, a fact closely related to the stronger dipolar interactions between the molecules of this amine. The negative $R_{\mathrm{m}}^{\mathrm{E}}$ values, which decrease with the length of the alkan-1-ol chain, indicate a loss of dispersive interactions and can be ascribed to the mentioned solvation effects. The Kirkwood-Fröhlich model supports the relevance of solvation in the dielectric properties of the mixtures, i.e., the presence linear-like multimers (especially for the methanol mixture) with dipole moments that respond effectively to an electric field and give a positive contribution to $\varepsilon_r^E$. Interestingly, the model suggests a less parallel alignment of molecular dipoles in the corresponding alkan-1-ol + aniline systems.

## Funding

This work was supported by Consejería de Educación de Castilla y León, under Project VA100G19 (Apoyo a GIR, BDNS: 425389). Ana Cobos gratefully acknowledges the support from Ministerio de Educación, Cultura y Deporte through the grant FPU15/05456.

Table 1

Sample description.

| Chemical name | CAS Number | Source | Purification method | Purity[a] | Water content[b] |
|---|---|---|---|---|---|
| methanol | 67-56-1 | Sigma-Aldrich | none | 0.999 | $2 \cdot 10^{-5}$ |
| propan-1-ol | 71-23-8 | Fluka | none | 0.999 | $1 \cdot 10^{-3}$ |
| pentan-1-ol | 71-41-0 | Sigma-Aldrich | none | 0.999 | $3 \cdot 10^{-4}$ |
| benzylamine | 100-46-9 | Fluka | none | 0.998 | $6.8 \cdot 10^{-4}$ |

[a] In mole fraction. By gas chromatography. Initial purity provided by the supplier.

[b] In mass fraction. By Karl-Fischer titration.

Table 2

Thermophysical properties of the pure liquids used in this work at temperature $T$ and pressure $p$ = 0.1 MPa: dipole moment ($\mu$), density ($\rho^*$), speed of sound ($c^*$), isobaric thermal expansion coefficient ($\alpha_p^*$), isentropic compressibility ($\kappa_S^*$), molar isobaric heat capacity ($C_{pm}^*$), isothermal compressibility ($\kappa_T^*$), refractive index at the sodium D-line ($n_D^*$) and relative permittivity at frequency $f$ = 1 MHz ($\varepsilon_r^*$). [a]

| Property | $T$/K | methanol | propan-1-ol | pentan-1-ol | benzylamine |
|---|---|---|---|---|---|
| $\mu$/D | | 1.664 [55] | 1.629 [55] | 1.598 [55] | 1.38 [56] |
| $\rho^*$/g·cm$^{-3}$ | 293.15 | 0.79191 | 0.80352 | 0.81454 | 0.98238 |
| | | 0.7916 [57] | 0.80361 [59] | 0.81468 [60] | 0.983 [61] |
| | | 0.791400 [58] | | | 0.98366 [29] |
| | | | | | 0.9834 [29] |
| | 298.15 | 0.78720 | 0.79951 | 0.81087 | 0.97809 |
| | | 0.7869 [62] | 0.79960 [59] | 0.81103 [60] | 0.978 [61] |
| | | 0.786884 [63] | | | 0.981 [26] |
| | | | | | 0.978337 [64] |
| | | | | | 0.97935 [29] |
| | | | | | 0.9789 [29] |
| | 303.15 | 0.78259 | 0.79547 | 0.80724 | 0.97370 |
| | | 0.782158 [63] | 0.79561 [59] | 0.81737 [60] | 0.9737 [65] |
| | | | | | 0.9737 [66] |
| | | | | | 0.9732[67] |
| | | | | | 0.974 [61] |
| | | | | | 0.97501 [29] |
| | | | | | 0.9744 [29] |
| | | | | | 0.9756 [27,28] |
| $c^*$/m·s$^{-1}$ | 293.15 | 1119.1 | 1222.5 | 1292.4 | 1580.1 |
| | | 1119 [68] | 1223 [69] | 1292 [68] | 1580.59 [29] |
| | | | | | 1579.7 [61] |
| | 298.15 | 1102.3 | 1205.1 | 1275.3 | 1559.9 |
| | | 1101.9 [70] | 1206 [69] | 1276 [68] | 1560.1 [61] |
| | | | | | 1561.06 [29] |
| | 303.15 | 1086.6 | 1188.6 | 1259.0 | 1539.9 |
| | | 1186.37 [71] | 1189 [69] | 1259 [68] | 1540.2 [61] |
| | | | | | 1541.20 [29] |
| | | | | | 1541.2 [66] |
| | | | | | 1538.2 [27,28] |
| $\alpha_p^*$/10$^{-3}$K$^{-1}$ | 298.15 | 1.184 | 1.007 | 0.900 | 0.887 |
| | | 1.196 [49] | 1.004 [49] | 0.905 [49] | 0.882 [29] |
| $\kappa_S^*$/TPa$^{-1}$ | 293.15 | 1008.3 | 832.7 | 735.0 | 407.7 |
| | | | | | 406.93 [29] |
| | 298.15 | 1045.5 | 861.3 | 758.3 | 420.2 |
| | | 1028 [49] | 849 [49] | 758 [72] | 419.01 [29] |
| | 303.15 | 1082.2 | 889.8 | 781.5 | 433.1 |
| | | | | | 431.79 [29] |
| $C_{pm}^*$/J·mol$^{-1}$·K$^{-1}$ | 298.15 | 81.92 [73] | 146.88 [73] | 207.45 [74] | 206.74 [64] |
| [b] $\kappa_T^*$/TPa$^{-1}$ | 298.15 | 1253.2 | 1016.0 | 884.9 | 544.5 |
| | | 1248 [49] | 1026 [49] | 884 [49] | |
| $n_D^*$ | 293.15 | 1.32852 | 1.38521 | 1.40989 | 1.54364 |
| | | 1.32859 [75] | 1.38512 [76] | 1.40986 [68] | 1.54380 [77] |
| | | | | | 1.544 [61] |
| | | | | | 1.5427 [29] |
| | | | | | 1.5445 [78] |
| | 298.15 | 1.32639 | 1.38305 | 1.40794 | 1.54132 |
| | | 1.32652 [79] | 1.38307 [68] | 1.40789 [68] | 1.5413 [80] |
| | | | | | 1.542 [61] |
| | | | | | 1.5387 [29] |
| | 303.15 | 1.32431 | 1.38103 | 1.40597 | 1.53893 |
| | | 1.32457 [81] | 1.38104 [68] | 1.40592 [82] | 1.539 [61] |
| | | 1.32410 [68] | | | 1.5393 [78] |
| | | | | | 1.5352 [29] |
| $\varepsilon_r^*$ | 293.15 | 33.576 | 21.224 | 15.745 | 5.019 |
| | | 33.61 [83] | 21.15 [84] | 15.63 [83] | 5.18 [85] |
| | 298.15 | 32.624 | 20.547 | 15.161 | 4.945 |
| | | 32.62 [83] | 20.42 [84] | 15.08 [86] | 5.20 [85] |
| | 303.15 | 31.684 | 19.875 | 14.580 | 4.871 |
| | | 31.66 [83] | 19.75 [84] | 14.44 [83] | |

[a] Standard uncertainties, $u$: $u(T) = 0.01$ K for $\rho^*$ and $c^*$ measurements; $u(T) = 0.02$ K for $\varepsilon_r^*$ and $n_D^*$ measurements; $u(p) = 10$ kPa; $u(f) = 20$ Hz. Expanded uncertainties, $U$ (coverage factor = 2, approx. 95% confidence level): $U(\rho^*) = 0.0006$ g·cm$^{-3}$, $U(c^*) = 1.3$ m·s$^{-1}$; $U(n_D^*) = 0.0003$. Relative expanded uncertainties, $U_r$ (coverage factor = 2): $U_r(\alpha_p^*) = 0.028$; $U_r(\kappa_S^*) = 0.002$; $U_r(\kappa_T^*) = 0.015$; $U_r(\varepsilon_r^*) = 0.01$.

[b] Determined using experimental values measured in this work and $C_{pm}^*$ values from the literature included in this table.

Table 3

Density ($\rho$), speed of sound ($c$) and excess molar volume ($V_m^E$) of alkan-1-ol (1) + benzylamine (2) liquid mixtures as functions of the mole fraction of the alkan-1-ol ($x_1$) at temperature $T$ and pressure $p$ = 0.1 MPa.[a]

| $x_1$ | $\rho$/g·cm⁻³ | $c$/m·s⁻¹ | $V_m^E$/cm³·mol⁻¹ | $x_1$ | $\rho$/g·cm⁻³ | $c$/m·s⁻¹ | $V_m^E$/cm³·mol⁻¹ |
|---|---|---|---|---|---|---|---|
| | | | | | | | |

$\rho$/g·cm⁻³, $c$/m·s⁻¹, $V_m^E$/cm³·mol⁻¹

| $x_1$ | $\rho$ | $c$ | $V_m^E$ | $x_1$ | $\rho$ | $c$ | $V_m^E$ |
|---|---|---|---|---|---|---|---|
| methanol (1) + benzylamine (2) ; $T$/K = 293.15 | | | | | | | |
| 0.0585 | 0.98040 | 1576.0 | -0.2477 | 0.5507 | 0.94342 | 1484.2 | -1.5547 |
| 0.1129 | 0.97838 | 1571.4 | -0.4751 | 0.6059 | 0.93479 | 1461.7 | -1.5588 |
| 0.1501 | 0.97677 | 1567.7 | -0.6170 | 0.6501 | 0.92655 | 1440.1 | -1.5227 |
| 0.2045 | 0.97418 | 1561.7 | -0.8178 | 0.7037 | 0.91501 | 1409.8 | -1.4511 |
| 0.2512 | 0.97161 | 1555.6 | -0.9745 | 0.7481 | 0.90354 | 1380.1 | -1.3423 |
| 0.3053 | 0.96823 | 1547.3 | -1.1422 | 0.8011 | 0.88761 | 1339.1 | -1.1785 |
| 0.3541 | 0.96459 | 1538.4 | -1.2656 | 0.8499 | 0.86982 | 1294.8 | -0.9616 |
| 0.4019 | 0.96060 | 1528.3 | -1.3766 | 0.9010 | 0.84806 | 1242.5 | -0.7035 |
| 0.4513 | 0.95573 | 1516.0 | -1.4611 | 0.9498 | 0.82297 | 1185.4 | -0.3898 |
| 0.4968 | 0.95063 | 1502.9 | -1.5231 | | | | |
| methanol (1) + benzylamine (2) ; $T$/K = 298.15 | | | | | | | |
| 0.0521 | 0.97648 | 1556.2 | -0.2399 | 0.5514 | 0.93872 | 1464.0 | -1.5616 |
| 0.1123 | 0.97421 | 1551.3 | -0.4901 | 0.6029 | 0.93078 | 1443.4 | -1.5744 |
| 0.1523 | 0.97252 | 1547.4 | -0.6494 | 0.6503 | 0.92202 | 1420.8 | -1.5407 |
| 0.1970 | 0.97037 | 1542.5 | -0.8138 | 0.7004 | 0.91113 | 1392.4 | -1.4656 |
| 0.2527 | 0.96733 | 1535.4 | -1.0047 | 0.7489 | 0.89868 | 1360.4 | -1.3507 |
| 0.2970 | 0.96454 | 1528.8 | -1.1408 | 0.7998 | 0.88328 | 1321.2 | -1.1870 |
| 0.3542 | 0.96029 | 1518.5 | -1.2873 | 0.8506 | 0.86497 | 1276.0 | -0.9725 |
| 0.3994 | 0.95677 | 1509.1 | -1.4151 | 0.9003 | 0.84366 | 1225.3 | -0.7121 |
| 0.4505 | 0.95143 | 1496.4 | -1.4784 | 0.9501 | 0.81807 | 1167.6 | -0.3900 |
| 0.4999 | 0.94581 | 1482.1 | -1.5424 | | | | |
| methanol (1) + benzylamine (2) ; $T$/K = 303.15 | | | | | | | |
| 0.0485 | 0.97209 | 1536.6 | -0.2137 | 0.5526 | 0.93411 | 1445.3 | -1.5834 |
| 0.1030 | 0.97010 | 1532.4 | -0.4488 | 0.6051 | 0.92581 | 1424.0 | -1.5847 |
| 0.1459 | 0.96831 | 1528.5 | -0.6226 | 0.6530 | 0.91692 | 1401.2 | -1.5523 |
| 0.1955 | 0.96589 | 1523.1 | -0.8032 | 0.7005 | 0.90658 | 1374.5 | -1.4825 |
| 0.2470 | 0.96308 | 1516.7 | -0.9815 | 0.7524 | 0.89322 | 1340.4 | -1.3605 |
| 0.2953 | 0.96007 | 1509.6 | -1.1341 | 0.8006 | 0.87850 | 1303.4 | -1.2008 |
| 0.3501 | 0.95614 | 1500.1 | -1.2884 | 0.8504 | 0.86055 | 1259.5 | -0.9897 |
| 0.3964 | 0.95215 | 1490.5 | -1.3869 | 0.9000 | 0.83923 | 1209.1 | -0.7237 |
| 0.4466 | 0.94730 | 1478.5 | -1.4822 | 0.9498 | 0.81374 | 1152.0 | -0.4027 |
| 0.4948 | 0.94186 | 1464.9 | -1.5464 | | | | |
| propan-1-ol (1) + benzylamine (2) ; $T$/K = 293.15 | | | | | | | |
| 0.0515 | 0.97739 | 1570.8 | -0.1570 | 0.5509 | 0.91138 | 1438.3 | -1.0601 |
| 0.1016 | 0.97239 | 1561.0 | -0.3129 | 0.6023 | 0.90196 | 1419.0 | -1.0487 |
| 0.1509 | 0.96713 | 1550.9 | -0.4489 | 0.6509 | 0.89245 | 1399.4 | -1.0140 |
| 0.2006 | 0.96160 | 1539.9 | -0.5821 | 0.7017 | 0.88184 | 1377.8 | -0.9515 |
| 0.2448 | 0.95634 | 1529.5 | -0.6828 | 0.7489 | 0.87133 | 1356.4 | -0.8683 |
| 0.3011 | 0.94930 | 1515.3 | -0.8018 | 0.7999 | 0.85925 | 1332.0 | -0.7526 |
| 0.3532 | 0.94233 | 1501.3 | -0.8928 | 0.8496 | 0.84671 | 1306.9 | -0.6127 |
| 0.4003 | 0.93563 | 1487.7 | -0.9589 | 0.9000 | 0.83311 | 1280.0 | -0.4364 |
| 0.4505 | 0.92807 | 1472.5 | -1.0143 | 0.9497 | 0.81885 | 1252.1 | -0.2341 |
| 0.5007 | 0.92000 | 1455.9 | -1.0485 | | | | |

propan-1-ol (1) + benzylamine (2) ; $T$/K = 298.15

| | | | | | | | |
|---|---|---|---|---|---|---|---|
| 0.0608 | 0.97211 | 1548.9 | -0.1785 | 0.5536 | 0.90653 | 1418.5 | -1.0521 |
| 0.1085 | 0.96730 | 1539.9 | -0.3252 | 0.5994 | 0.89805 | 1401.0 | -1.0322 |
| 0.1554 | 0.96228 | 1530.1 | -0.4556 | 0.6517 | 0.88793 | 1380.5 | -1.0035 |
| 0.2047 | 0.95670 | 1519.2 | -0.5801 | 0.7000 | 0.87783 | 1359.8 | -0.9407 |
| 0.2518 | 0.95112 | 1508.3 | -0.6927 | 0.7505 | 0.86662 | 1337.2 | -0.8527 |
| 0.3023 | 0.94471 | 1495.4 | -0.7917 | 0.7985 | 0.85533 | 1314.4 | -0.7487 |
| 0.3469 | 0.93881 | 1483.6 | -0.8754 | 0.8425 | 0.84431 | 1292.4 | -0.6263 |
| 0.3980 | 0.93153 | 1469.1 | -0.9435 | 0.8987 | 0.82931 | 1262.7 | -0.4364 |
| 0.4446 | 0.92457 | 1455.0 | -0.9982 | 0.9499 | 0.81470 | 1234.4 | -0.2305 |
| 0.4912 | 0.91706 | 1439.8 | -1.0232 | | | | |

propan-1-ol (1) + benzylamine (2) ; $T$/K = 303.15

| | | | | | | | |
|---|---|---|---|---|---|---|---|
| 0.0511 | 0.96877 | 1530.9 | -0.1592 | 0.5506 | 0.90270 | 1400.9 | -1.0514 |
| 0.1010 | 0.96377 | 1521.5 | -0.3131 | 0.6007 | 0.89358 | 1382.5 | -1.0428 |
| 0.1530 | 0.95823 | 1510.9 | -0.4585 | 0.6498 | 0.88398 | 1363.0 | -1.0039 |
| 0.2000 | 0.95296 | 1500.7 | -0.5810 | 0.6989 | 0.87379 | 1342.5 | -0.9439 |
| 0.2525 | 0.94668 | 1488.6 | -0.6991 | 0.7468 | 0.86325 | 1321.3 | -0.8648 |
| 0.3042 | 0.94018 | 1475.7 | -0.8073 | 0.8014 | 0.85032 | 1295.4 | -0.7342 |
| 0.3476 | 0.93433 | 1464.3 | -0.8776 | 0.8502 | 0.83805 | 1271.1 | -0.5951 |
| 0.4168 | 0.92440 | 1444.5 | -0.9690 | 0.9005 | 0.82464 | 1244.8 | -0.4272 |
| 0.4517 | 0.91909 | 1433.9 | -1.0047 | 0.9501 | 0.81055 | 1217.4 | -0.2301 |
| 0.5020 | 0.91104 | 1417.8 | -1.0417 | | | | |

pentan-1-ol (1) + benzylamine (2) ; $T$/K = 293.15

| | | | | | | | |
|---|---|---|---|---|---|---|---|
| 0.0513 | 0.97507 | 1567.1 | -0.1382 | 0.5485 | 0.89705 | 1429.6 | -0.7753 |
| 0.0974 | 0.96834 | 1555.2 | -0.2467 | 0.5991 | 0.88834 | 1414.6 | -0.7572 |
| 0.1492 | 0.96063 | 1541.5 | -0.3544 | 0.6480 | 0.87980 | 1399.8 | -0.7252 |
| 0.1986 | 0.95321 | 1528.2 | -0.4517 | 0.6982 | 0.87090 | 1384.9 | -0.6760 |
| 0.2499 | 0.94539 | 1514.5 | -0.5419 | 0.7498 | 0.86156 | 1369.2 | -0.6014 |
| 0.2978 | 0.93795 | 1501.3 | -0.6123 | 0.7986 | 0.85262 | 1354.3 | -0.5167 |
| 0.3478 | 0.93004 | 1487.3 | -0.6709 | 0.8484 | 0.84335 | 1339.1 | -0.4105 |
| 0.3999 | 0.92172 | 1472.6 | -0.7247 | 0.8991 | 0.83380 | 1323.4 | -0.2864 |
| 0.4490 | 0.91372 | 1458.6 | -0.7582 | 0.9496 | 0.82415 | 1307.7 | -0.1429 |
| 0.4961 | 0.90589 | 1444.9 | -0.7731 | | | | |

pentan-1-ol (1) + benzylamine (2) ; $T$/K = 298.15

| | | | | | | | |
|---|---|---|---|---|---|---|---|
| 0.0506 | 0.97084 | 1547.4 | -0.1297 | 0.5504 | 0.89259 | 1410.9 | -0.7600 |
| 0.1001 | 0.96358 | 1534.7 | -0.2400 | 0.6009 | 0.88402 | 1396.0 | -0.7528 |
| 0.1477 | 0.95658 | 1522.5 | -0.3462 | 0.6422 | 0.87679 | 1383.8 | -0.7201 |
| 0.1983 | 0.94898 | 1509.1 | -0.4432 | 0.6987 | 0.86684 | 1367.0 | -0.6680 |
| 0.2486 | 0.94133 | 1495.7 | -0.5309 | 0.7496 | 0.85759 | 1351.3 | -0.5849 |
| 0.2968 | 0.93380 | 1482.5 | -0.5937 | 0.8005 | 0.84841 | 1336.2 | -0.5099 |

| $x_1$ | $\rho$ | $c$ | $V_m^E$ | $x_1$ | $\rho$ | $c$ | $V_m^E$ |
|---|---|---|---|---|---|---|---|
| 0.3476 | 0.92588 | 1468.7 | -0.6636 | 0.8492 | 0.83929 | 1321.3 | -0.3939 |
| 0.3987 | 0.91773 | 1454.4 | -0.7142 | 0.9002 | 0.82983 | 1305.8 | -0.2825 |
| 0.4501 | 0.90938 | 1439.9 | -0.7487 | 0.9500 | 0.82034 | 1290.5 | -0.1389 |
| 0.4985 | 0.90135 | 1425.9 | -0.7624 | | | | |

<div align="center">pentan-1-ol (1) + benzylamine (2) ; $T/K = 303.15$</div>

| $x_1$ | $\rho$ | $c$ | $V_m^E$ | $x_1$ | $\rho$ | $c$ | $V_m^E$ |
|---|---|---|---|---|---|---|---|
| 0.0482 | 0.96687 | 1528.2 | -0.1291 | 0.5527 | 0.88816 | 1392.5 | -0.7580 |
| 0.1009 | 0.95917 | 1514.9 | -0.2464 | 0.6037 | 0.87938 | 1377.6 | -0.7304 |
| 0.1497 | 0.95197 | 1502.4 | -0.3494 | 0.6497 | 0.87145 | 1364.1 | -0.7044 |
| 0.1995 | 0.94451 | 1489.6 | -0.4437 | 0.7039 | 0.86188 | 1348.1 | -0.6454 |
| 0.2494 | 0.93693 | 1476.6 | -0.5281 | 0.7504 | 0.85356 | 1334.3 | -0.5807 |
| 0.2989 | 0.92928 | 1463.3 | -0.5985 | 0.8012 | 0.84424 | 1318.8 | -0.4796 |
| 0.3521 | 0.92097 | 1448.9 | -0.6658 | 0.8505 | 0.83512 | 1303.9 | -0.3705 |
| 0.4003 | 0.91326 | 1435.6 | -0.7070 | 0.9008 | 0.82580 | 1288.8 | -0.2556 |
| 0.4529 | 0.90475 | 1420.9 | -0.7419 | 0.9506 | 0.81650 | 1273.9 | -0.1306 |
| 0.5006 | 0.89685 | 1407.4 | -0.7527 | | | | |

[a] Standard uncertainties, $u$: $u(T) = 0.01$ K; $u(p) = 10$ kPa. Expanded uncertainties, $U$ (coverage factor = 2, approx. 95% confidence level): $U(x_1) = 0.0010$; $U(\rho) = 0.0006$ g·cm$^{-3}$; $U(c) = 1.3$ m·s$^{-1}$; $U(V_m^E) = 0.010 \cdot |V_m^E|_{max} + 0.005$ cm$^3$·mol$^{-1}$.

Table 4

Excess isentropic compressibility ($\kappa_S^E$) and excess speed of sound ($c^E$) of alkan-1-ol (1) + benzylamine (2) liquid mixtures as functions of the mole fraction of the alkan-1-ol ($x_1$) at temperature $T = 298.15$ K and pressure $p = 0.1$ MPa.[a]

| $x_1$ | $\kappa_S^E$/TPa$^{-1}$ | $c^E$/m·s$^{-1}$ | $x_1$ | $\kappa_S^E$/TPa$^{-1}$ | $c^E$/m·s$^{-1}$ |
|---|---|---|---|---|---|
| | | methanol (1) + benzylamine (2) ; $T$/K = 298.15 | | | |
| 0.0521 | -10.0 | 16.3 | 0.5514 | -121.0 | 136.5 |
| 0.1123 | -22.1 | 35.0 | 0.6029 | -132.0 | 140.3 |
| 0.1523 | -30.4 | 47.1 | 0.6503 | -140.4 | 140.4 |
| 0.1970 | -39.9 | 60.2 | 0.7004 | -146.8 | 136.5 |
| 0.2527 | -52.3 | 76.1 | 0.7489 | -149.6 | 128.6 |
| 0.2970 | -62.4 | 88.2 | 0.7998 | -147.2 | 115.2 |
| 0.3542 | -75.6 | 102.4 | 0.8506 | -136.5 | 96.1 |
| 0.3994 | -86.5 | 112.7 | 0.9003 | -113.9 | 71.2 |
| 0.4505 | -98.3 | 122.7 | 0.9501 | -72.3 | 39.4 |
| 0.4999 | -109.8 | 130.8 | | | |
| | | propan-1-ol (1) + benzylamine (2) ; $T$/K = 298.15 | | | |
| 0.0608 | -10.2 | 16.8 | 0.5536 | -75.0 | 80.3 |
| 0.1085 | -18.3 | 29.1 | 0.5994 | -76.6 | 78.3 |
| 0.1554 | -25.9 | 39.5 | 0.6517 | -77.5 | 74.9 |
| 0.2047 | -33.7 | 49.3 | 0.7000 | -75.9 | 69.6 |
| 0.2518 | -41.0 | 57.7 | 0.7505 | -72.2 | 62.6 |
| 0.3023 | -48.2 | 65.0 | 0.7985 | -66.2 | 54.2 |
| 0.3469 | -54.3 | 70.5 | 0.8425 | -57.9 | 44.9 |
| 0.3980 | -60.7 | 75.3 | 0.8987 | -42.9 | 30.9 |
| 0.4446 | -65.9 | 78.4 | 0.9499 | -24.3 | 16.4 |
| 0.4912 | -70.2 | 79.9 | | | |
| | | pentan-1-ol (1) + benzylamine (2) ; $T$/K = 298.15 | | | |
| 0.0506 | -7.0 | 11.5 | 0.5504 | -42.9 | 46.1 |
| 0.1001 | -13.2 | 20.8 | 0.6009 | -42.3 | 43.6 |
| 0.1477 | -18.8 | 28.4 | 0.6422 | -41.1 | 41.0 |
| 0.1983 | -24.1 | 34.9 | 0.6987 | -38.6 | 36.7 |
| 0.2486 | -28.9 | 40.0 | 0.7496 | -34.6 | 31.6 |
| 0.2968 | -32.8 | 43.6 | 0.8005 | -30.3 | 26.6 |
| 0.3476 | -36.5 | 46.4 | 0.8492 | -24.5 | 20.8 |
| 0.3987 | -39.3 | 47.9 | 0.9002 | -17.6 | 14.3 |
| 0.4501 | -41.4 | 48.3 | 0.9500 | -9.3 | 7.3 |
| 0.4985 | -42.5 | 47.6 | | | |

[a] Standard uncertainties, $u$: $u(T) = 0.01$ K; $u(p) = 10$ kPa. Expanded uncertainty, $U$ (coverage factor = 2, approx. 95% confidence level): $U(x_1) = 0.0010$. Relative expanded uncertainties, $U_r$: $U_r(\kappa_S^E) = 0.015$, $U_r(c^E) = 0.02$.

Table 5

Volume fractions of alkan-1-ol ($\phi_1$), relative permittivities at frequency $f = 1$ MHz ($\varepsilon_r$) and excess relative permittivities at $f = 1$ MHz ($\varepsilon_r^E$) of alkan-1-ol (1) + benzylamine (2) liquid mixtures as functions of the mole fraction of the alkan-1-ol ($x_1$), at temperature $T$ and pressure $p$ = 0.1 MPa. [a]

| $x_1$ | $\phi_1$ | $\varepsilon_r$ | $\varepsilon_r^E$ | $x_1$ | $\phi_1$ | $\varepsilon_r$ | $\varepsilon_r^E$ |
|---|---|---|---|---|---|---|---|
| | | | methanol (1) + benzylamine (2) ; $T$/K = 293.15 | | | | |
| 0.0491 | 0.0188 | 5.490 | -0.066 | 0.5528 | 0.3144 | 14.111 | 0.114 |
| 0.1123 | 0.0448 | 6.119 | -0.179 | 0.5952 | 0.3529 | 15.326 | 0.229 |
| 0.1520 | 0.0623 | 6.580 | -0.218 | 0.6515 | 0.4095 | 17.197 | 0.484 |
| 0.2044 | 0.0870 | 7.216 | -0.287 | 0.6944 | 0.4574 | 18.637 | 0.556 |
| 0.2499 | 0.1100 | 7.864 | -0.296 | 0.7498 | 0.5264 | 20.808 | 0.757 |
| 0.2989 | 0.1366 | 8.595 | -0.325 | 0.7983 | 0.5948 | 22.808 | 0.803 |
| 0.3617 | 0.1737 | 9.709 | -0.270 | 0.8516 | 0.6804 | 25.256 | 0.807 |
| 0.3922 | 0.1931 | 10.264 | -0.269 | 0.8991 | 0.7677 | 27.653 | 0.711 |
| 0.4506 | 0.2333 | 11.533 | -0.148 | 0.9500 | 0.8757 | 30.509 | 0.483 |
| 0.4988 | 0.2696 | 12.652 | -0.066 | | | | |
| | | | methanol (1) + benzylamine (2) ; $T$/K = 298.15 | | | | |
| 0.0491 | 0.0188 | 5.401 | -0.064 | 0.5528 | 0.3148 | 13.740 | 0.082 |
| 0.1123 | 0.0449 | 6.011 | -0.177 | 0.5952 | 0.3534 | 14.913 | 0.186 |
| 0.1520 | 0.0625 | 6.455 | -0.220 | 0.6515 | 0.4100 | 16.725 | 0.432 |
| 0.2044 | 0.0872 | 7.069 | -0.290 | 0.6944 | 0.4579 | 18.119 | 0.500 |
| 0.2499 | 0.1102 | 7.697 | -0.298 | 0.7498 | 0.5270 | 20.241 | 0.709 |
| 0.2989 | 0.1368 | 8.405 | -0.326 | 0.7983 | 0.5953 | 22.166 | 0.744 |
| 0.3617 | 0.1740 | 9.480 | -0.281 | 0.8516 | 0.6808 | 24.545 | 0.756 |
| 0.3922 | 0.1935 | 10.019 | -0.282 | 0.8991 | 0.7681 | 26.872 | 0.667 |
| 0.4506 | 0.2336 | 11.243 | -0.168 | 0.9500 | 0.8760 | 29.659 | 0.467 |
| 0.4988 | 0.2700 | 12.329 | -0.089 | | | | |
| | | | methanol (1) + benzylamine (2) ; $T$/K = 303.15 | | | | |
| 0.0491 | 0.0188 | 5.312 | -0.063 | 0.5528 | 0.3150 | 13.375 | 0.058 |
| 0.1123 | 0.0450 | 5.903 | -0.175 | 0.5952 | 0.3536 | 14.506 | 0.154 |
| 0.1520 | 0.0625 | 6.333 | -0.214 | 0.6515 | 0.4102 | 16.262 | 0.392 |
| 0.2044 | 0.0872 | 6.924 | -0.285 | 0.6944 | 0.4581 | 17.608 | 0.454 |
| 0.2499 | 0.1103 | 7.534 | -0.294 | 0.7498 | 0.5272 | 19.658 | 0.651 |
| 0.2989 | 0.1369 | 8.216 | -0.326 | 0.7983 | 0.5956 | 21.533 | 0.692 |
| 0.3617 | 0.1741 | 9.256 | -0.283 | 0.8516 | 0.6810 | 23.844 | 0.713 |
| 0.3922 | 0.1936 | 9.775 | -0.287 | 0.8991 | 0.7683 | 26.096 | 0.625 |
| 0.4506 | 0.2338 | 10.961 | -0.179 | 0.9500 | 0.8761 | 28.824 | 0.462 |
| 0.4988 | 0.2702 | 12.006 | -0.110 | | | | |
| | | | propan-1-ol (1) + benzylamine (2) ; $T$/K = 293.15 | | | | |
| 0.0515 | 0.0359 | 5.383 | -0.218 | 0.5509 | 0.4569 | 10.880 | -1.543 |
| 0.1016 | 0.0720 | 5.765 | -0.421 | 0.6023 | 0.5094 | 11.736 | -1.538 |
| 0.1494 | 0.1075 | 6.159 | -0.602 | 0.6509 | 0.5611 | 12.629 | -1.483 |
| 0.2006 | 0.1468 | 6.610 | -0.788 | 0.7017 | 0.6173 | 13.631 | -1.391 |
| 0.2448 | 0.1818 | 7.042 | -0.923 | 0.7489 | 0.6716 | 14.646 | -1.256 |
| 0.3011 | 0.2280 | 7.597 | -1.117 | 0.7999 | 0.7327 | 15.833 | -1.059 |
| 0.3532 | 0.2724 | 8.178 | -1.255 | 0.8496 | 0.7948 | 17.054 | -0.845 |
| 0.4003 | 0.3140 | 8.742 | -1.365 | 0.9000 | 0.8606 | 18.390 | -0.575 |
| 0.4505 | 0.3599 | 9.398 | -1.453 | 0.9497 | 0.9283 | 19.774 | -0.288 |
| 0.5007 | 0.4074 | 10.107 | -1.514 | | | | |

propan-1-ol (1) + benzylamine (2) ; $T$/K = 298.15

| | | | | | | | |
|---|---|---|---|---|---|---|---|
| 0.0515 | 0.0359 | 5.295 | -0.210 | 0.5509 | 0.4570 | 10.576 | -1.499 |
| 0.1016 | 0.0720 | 5.663 | -0.405 | 0.6023 | 0.5096 | 11.392 | -1.504 |
| 0.1494 | 0.1076 | 6.045 | -0.579 | 0.6509 | 0.5613 | 12.251 | -1.451 |
| 0.2006 | 0.1469 | 6.477 | -0.760 | 0.7017 | 0.6175 | 13.206 | -1.373 |
| 0.2448 | 0.1820 | 6.897 | -0.888 | 0.7489 | 0.6718 | 14.185 | -1.241 |
| 0.3011 | 0.2282 | 7.427 | -1.078 | 0.7999 | 0.7328 | 15.322 | -1.056 |
| 0.3532 | 0.2726 | 7.986 | -1.212 | 0.8496 | 0.7949 | 16.502 | -0.845 |
| 0.4003 | 0.3141 | 8.525 | -1.321 | 0.9000 | 0.8606 | 17.789 | -0.583 |
| 0.4505 | 0.3600 | 9.156 | -1.406 | 0.9497 | 0.9283 | 19.138 | -0.290 |
| 0.5007 | 0.4076 | 9.832 | -1.472 | | | | |

propan-1-ol (1) + benzylamine (2) ; $T$/K = 303.15

| | | | | | | | |
|---|---|---|---|---|---|---|---|
| 0.0515 | 0.0359 | 5.211 | -0.199 | 0.5509 | 0.4571 | 10.277 | -1.452 |
| 0.1016 | 0.0720 | 5.563 | -0.388 | 0.6023 | 0.5097 | 11.056 | -1.463 |
| 0.1494 | 0.1076 | 5.932 | -0.553 | 0.6509 | 0.5614 | 11.880 | -1.414 |
| 0.2006 | 0.1470 | 6.345 | -0.732 | 0.7017 | 0.6176 | 12.790 | -1.347 |
| 0.2448 | 0.1820 | 6.756 | -0.846 | 0.7489 | 0.6719 | 13.733 | -1.219 |
| 0.3011 | 0.2283 | 7.259 | -1.037 | 0.7999 | 0.7329 | 14.821 | -1.046 |
| 0.3532 | 0.2727 | 7.796 | -1.167 | 0.8496 | 0.7950 | 15.957 | -0.842 |
| 0.4003 | 0.3142 | 8.310 | -1.275 | 0.9000 | 0.8607 | 17.198 | -0.587 |
| 0.4505 | 0.3601 | 8.918 | -1.356 | 0.9497 | 0.9284 | 18.508 | -0.293 |
| 0.5007 | 0.4077 | 9.562 | -1.426 | | | | |

pentan-1-ol (1) + benzylamine (2) ; $T$/K = 293.15

| | | | | | | | |
|---|---|---|---|---|---|---|---|
| 0.0513 | 0.0509 | 5.311 | -0.254 | 0.5485 | 0.5466 | 8.830 | -2.052 |
| 0.0974 | 0.0967 | 5.590 | -0.466 | 0.5991 | 0.5972 | 9.335 | -2.090 |
| 0.1492 | 0.1482 | 5.898 | -0.711 | 0.6480 | 0.6462 | 9.860 | -2.090 |
| 0.1986 | 0.1974 | 6.208 | -0.928 | 0.6982 | 0.6965 | 10.475 | -2.015 |
| 0.2499 | 0.2484 | 6.536 | -1.147 | 0.7498 | 0.7483 | 11.164 | -1.881 |
| 0.2978 | 0.2962 | 6.856 | -1.340 | 0.7986 | 0.7973 | 11.908 | -1.663 |
| 0.3478 | 0.3460 | 7.203 | -1.527 | 0.8484 | 0.8474 | 12.727 | -1.381 |
| 0.3999 | 0.3980 | 7.588 | -1.700 | 0.8991 | 0.8984 | 13.672 | -0.983 |
| 0.4490 | 0.4471 | 7.966 | -1.849 | 0.9496 | 0.9492 | 14.675 | -0.525 |
| 0.4961 | 0.4941 | 8.367 | -1.952 | | | | |

pentan-1-ol (1) + benzylamine (2) ; $T$/K = 298.15

| | | | | | | | |
|---|---|---|---|---|---|---|---|
| 0.0513 | 0.0509 | 5.226 | -0.239 | 0.5485 | 0.5466 | 8.596 | -1.933 |
| 0.0974 | 0.0967 | 5.495 | -0.438 | 0.5991 | 0.5973 | 9.072 | -1.975 |
| 0.1492 | 0.1482 | 5.791 | -0.668 | 0.6480 | 0.6463 | 9.572 | -1.976 |
| 0.1986 | 0.1974 | 6.089 | -0.873 | 0.6982 | 0.6966 | 10.150 | -1.911 |
| 0.2499 | 0.2485 | 6.404 | -1.080 | 0.7498 | 0.7484 | 10.805 | -1.786 |
| 0.2978 | 0.2962 | 6.712 | -1.259 | 0.7986 | 0.7974 | 11.501 | -1.590 |
| 0.3478 | 0.3461 | 7.045 | -1.436 | 0.8484 | 0.8474 | 12.283 | -1.319 |
| 0.3999 | 0.3981 | 7.412 | -1.600 | 0.8991 | 0.8984 | 13.173 | -0.950 |
| 0.4490 | 0.4471 | 7.774 | -1.739 | 0.9496 | 0.9492 | 14.135 | -0.507 |
| 0.4961 | 0.4942 | 8.153 | -1.841 | | | | |
| pentan-1-ol (1) + benzylamine (2) ; $T/K = 303.15$ | | | | | | | |
| 0.0513 | 0.0509 | 5.142 | -0.223 | 0.5485 | 0.5466 | 8.366 | -1.812 |
| 0.0974 | 0.0967 | 5.400 | -0.410 | 0.5991 | 0.5972 | 8.813 | -1.856 |
| 0.1492 | 0.1482 | 5.687 | -0.623 | 0.6480 | 0.6462 | 9.289 | -1.856 |
| 0.1986 | 0.1974 | 5.971 | -0.817 | 0.6982 | 0.6966 | 9.832 | -1.802 |
| 0.2499 | 0.2485 | 6.275 | -1.009 | 0.7498 | 0.7483 | 10.451 | -1.685 |
| 0.2978 | 0.2962 | 6.568 | -1.179 | 0.7986 | 0.7974 | 11.104 | -1.509 |
| 0.3478 | 0.3460 | 6.889 | -1.341 | 0.8484 | 0.8474 | 11.845 | -1.253 |
| 0.3999 | 0.3980 | 7.238 | -1.497 | 0.8991 | 0.8984 | 12.682 | -0.912 |
| 0.4490 | 0.4471 | 7.585 | -1.627 | 0.9496 | 0.9492 | 13.603 | -0.484 |
| 0.4961 | 0.4942 | 7.943 | -1.726 | | | | |

Table 6

Volume fractions of alkan-1-ol ($\phi_1$), refractive indices at the sodium D-line ($n_D$) and excess refractive indices at the sodium D-line ($n_D^E$) of alkan-1-ol (1) + benzylamine (2) liquid mixtures as functions of the mole fraction of the alkan-1-ol, $x_1$, at temperature $T$ and pressure $p = 0.1$ MPa. [a]

| $x_1$ | $\phi_1$ | $n_D$ | $10^5 n_D^E$ | $x_1$ | $\phi_1$ | $n_D$ | $10^5 n_D^E$ |
|---|---|---|---|---|---|---|---|
| \multicolumn{8}{c}{methanol (1) + benzylamine (2) ; $T/K = 293.15$} |
| 0.0491 | 0.0188 | 1.54122 | 135 | 0.5528 | 0.3144 | 1.48728 | 790 |
| 0.1520 | 0.0623 | 1.53423 | 311 | 0.6515 | 0.4095 | 1.46711 | 772 |
| 0.2499 | 0.1100 | 1.52608 | 461 | 0.7498 | 0.5264 | 1.44126 | 683 |
| 0.3617 | 0.1737 | 1.51461 | 613 | 0.8516 | 0.6804 | 1.40590 | 503 |
| 0.4506 | 0.2333 | 1.50312 | 690 | 0.9500 | 0.8757 | 1.35930 | 218 |
| \multicolumn{8}{c}{methanol (1) + benzylamine (2) ; $T/K = 298.15$} |
| 0.0491 | 0.0188 | 1.53772 | 16 | 0.6515 | 0.4100 | 1.46470 | 766 |
| 0.1520 | 0.0625 | 1.53077 | 200 | 0.7498 | 0.5270 | 1.43893 | 685 |
| 0.2499 | 0.1102 | 1.52268 | 355 | 0.8516 | 0.6808 | 1.40271 | 512 |
| 0.3617 | 0.1740 | 1.51140 | 527 | 0.8991 | 0.7681 | 1.38261 | 339 |
| 0.4506 | 0.2336 | 1.50041 | 653 | 0.9500 | 0.8760 | 1.35672 | 183 |
| 0.5528 | 0.3148 | 1.48450 | 746 | | | | |
| \multicolumn{8}{c}{methanol (1) + benzylamine (2) ; $T/K = 303.15$} |
| 0.1123 | 0.0450 | 1.53158 | 166 | 0.4506 | 0.2338 | 1.49800 | 648 |
| 0.1520 | 0.0625 | 1.52900 | 260 | 0.5528 | 0.3150 | 1.48200 | 730 |
| 0.2044 | 0.0872 | 1.52487 | 345 | 0.6515 | 0.4102 | 1.46200 | 727 |
| 0.2989 | 0.1369 | 1.51643 | 508 | 0.7498 | 0.5272 | 1.43600 | 620 |
| 0.3922 | 0.1936 | 1.50582 | 604 | 0.8991 | 0.7683 | 1.38083 | 381 |
| \multicolumn{8}{c}{propan-1-ol (1) + benzylamine (2) ; $T/K = 293.15$} |
| 0.0515 | 0.0359 | 1.53877 | 54 | 0.5509 | 0.4569 | 1.47666 | 329 |
| 0.1016 | 0.0720 | 1.53379 | 101 | 0.6023 | 0.5094 | 1.46835 | 327 |
| 0.1494 | 0.1075 | 1.52878 | 138 | 0.6509 | 0.5611 | 1.45996 | 309 |
| 0.2006 | 0.1468 | 1.52323 | 181 | 0.7017 | 0.6173 | 1.45086 | 297 |
| 0.2448 | 0.1818 | 1.51817 | 210 | 0.7489 | 0.6716 | 1.44175 | 259 |
| 0.3011 | 0.2280 | 1.51147 | 249 | 0.7999 | 0.7327 | 1.43199 | 271 |
| 0.3532 | 0.2724 | 1.50485 | 271 | 0.8496 | 0.7948 | 1.42108 | 192 |
| 0.4003 | 0.3140 | 1.49884 | 314 | 0.9000 | 0.8606 | 1.40987 | 151 |
| 0.4505 | 0.3599 | 1.49182 | 326 | 0.9497 | 0.9283 | 1.39810 | 93 |
| 0.5007 | 0.4074 | 1.48464 | 350 | | | | |
| \multicolumn{8}{c}{propan-1-ol (1) + benzylamine (2) ; $T/K = 298.15$} |
| 0.0515 | 0.0359 | 1.53644 | 52 | 0.5509 | 0.4570 | 1.47431 | 321 |
| 0.1016 | 0.0720 | 1.53154 | 107 | 0.6023 | 0.5096 | 1.46591 | 310 |
| 0.1494 | 0.1076 | 1.52648 | 140 | 0.6509 | 0.5613 | 1.45760 | 300 |
| 0.2006 | 0.1469 | 1.52096 | 186 | 0.7017 | 0.6175 | 1.44842 | 278 |
| 0.2448 | 0.1820 | 1.51594 | 219 | 0.7489 | 0.6718 | 1.43937 | 245 |
| 0.3011 | 0.2282 | 1.50936 | 269 | 0.7999 | 0.7328 | 1.42924 | 218 |

| 0.3532 | 0.2726 | 1.50256 | 273 | 0.8496 | 0.7949 | 1.41852 | 157 |
|--------|--------|---------|-----|--------|--------|---------|-----|
| 0.4003 | 0.3141 | 1.49647 | 305 | 0.9000 | 0.8606 | 1.40736 | 118 |
| 0.4505 | 0.3600 | 1.48942 | 313 | 0.9497 | 0.9283 | 1.39550 | 50 |
| 0.5007 | 0.4076 | 1.48216 | 330 | | | | |

<div align="center">propan-1-ol (1) + benzylamine (2) ; <em>T</em>/K = 303.15</div>

| 0.0515 | 0.0359 | 1.53406 | 52 | 0.5509 | 0.4571 | 1.47207 | 321 |
|--------|--------|---------|-----|--------|--------|---------|-----|
| 0.1016 | 0.0720 | 1.52905 | 94 | 0.6023 | 0.5097 | 1.46371 | 313 |
| 0.1494 | 0.1076 | 1.52411 | 138 | 0.6509 | 0.5614 | 1.45539 | 299 |
| 0.2006 | 0.1470 | 1.51853 | 178 | 0.7017 | 0.6176 | 1.44631 | 286 |
| 0.2448 | 0.1820 | 1.51356 | 214 | 0.7489 | 0.6719 | 1.43718 | 243 |
| 0.3011 | 0.2283 | 1.50685 | 251 | 0.7999 | 0.7329 | 1.42723 | 231 |
| 0.3532 | 0.2727 | 1.50033 | 281 | 0.8496 | 0.7950 | 1.41628 | 144 |
| 0.4003 | 0.3142 | 1.49410 | 298 | 0.9000 | 0.8607 | 1.40520 | 111 |
| 0.4505 | 0.3601 | 1.48714 | 313 | 0.9497 | 0.9284 | 1.39349 | 56 |
| 0.5007 | 0.4077 | 1.47980 | 321 | | | | |

<div align="center">pentan-1-ol (1) + benzylamine (2) ; <em>T</em>/K = 293.15</div>

| 0.0513 | 0.0509 | 1.53766 | 55 | 0.5485 | 0.5466 | 1.47359 | 155 |
|--------|--------|---------|-----|--------|--------|---------|-----|
| 0.0974 | 0.0967 | 1.53190 | 68 | 0.5991 | 0.5972 | 1.46663 | 140 |
| 0.1492 | 0.1482 | 1.52549 | 93 | 0.6480 | 0.6462 | 1.45989 | 128 |
| 0.1986 | 0.1974 | 1.51923 | 106 | 0.6982 | 0.6965 | 1.45288 | 109 |
| 0.2499 | 0.2484 | 1.51275 | 123 | 0.7498 | 0.7483 | 1.44573 | 101 |
| 0.2978 | 0.2962 | 1.50647 | 121 | 0.7986 | 0.7973 | 1.43869 | 68 |
| 0.3478 | 0.3460 | 1.50029 | 158 | 0.8484 | 0.8474 | 1.43171 | 60 |
| 0.3999 | 0.3980 | 1.49337 | 153 | 0.8991 | 0.8984 | 1.42443 | 38 |
| 0.4490 | 0.4471 | 1.48692 | 159 | 0.9496 | 0.9492 | 1.41706 | 7 |
| 0.4961 | 0.4941 | 1.48051 | 144 | | | | |

<div align="center">pentan-1-ol (1) + benzylamine (2) ; <em>T</em>/K = 298.15</div>

| 0.0513 | 0.0509 | 1.53519 | 38 | 0.5485 | 0.5466 | 1.47134 | 143 |
|--------|--------|---------|-----|--------|--------|---------|-----|
| 0.0974 | 0.0967 | 1.52951 | 58 | 0.5991 | 0.5973 | 1.46439 | 127 |
| 0.1492 | 0.1482 | 1.52301 | 72 | 0.6480 | 0.6463 | 1.45779 | 128 |
| 0.1986 | 0.1974 | 1.51688 | 96 | 0.6982 | 0.6966 | 1.45077 | 107 |
| 0.2499 | 0.2485 | 1.51041 | 113 | 0.7498 | 0.7484 | 1.44361 | 95 |
| 0.2978 | 0.2962 | 1.50410 | 105 | 0.7986 | 0.7974 | 1.43664 | 68 |
| 0.3478 | 0.3461 | 1.49783 | 133 | 0.8484 | 0.8474 | 1.42971 | 61 |
| 0.3999 | 0.3981 | 1.49092 | 127 | 0.8991 | 0.8984 | 1.42245 | 39 |
| 0.4490 | 0.4471 | 1.48458 | 141 | 0.9496 | 0.9492 | 1.41515 | 13 |
| 0.4961 | 0.4942 | 1.47828 | 137 | | | | |

<div align="center">pentan-1-ol (1) + benzylamine (2) ; <em>T</em>/K = 303.15</div>

| 0.0513 | 0.0509 | 1.53273 | 29 | 0.5485 | 0.5466 | 1.46909 | 134 |
|--------|--------|---------|-----|--------|--------|---------|-----|
| 0.0974 | 0.0967 | 1.52701 | 43 | 0.5991 | 0.5972 | 1.46225 | 127 |
| 0.1492 | 0.1482 | 1.52062 | 66 | 0.6480 | 0.6462 | 1.45555 | 115 |
| 0.1986 | 0.1974 | 1.51444 | 83 | 0.6982 | 0.6966 | 1.44867 | 107 |
| 0.2499 | 0.2485 | 1.50805 | 106 | 0.7498 | 0.7483 | 1.44142 | 83 |
| 0.2978 | 0.2962 | 1.50179 | 101 | 0.7986 | 0.7974 | 1.43459 | 69 |
| 0.3478 | 0.3460 | 1.49552 | 125 | 0.8484 | 0.8474 | 1.42770 | 64 |
| 0.3999 | 0.3980 | 1.48874 | 130 | 0.8991 | 0.8984 | 1.42048 | 43 |
| 0.4490 | 0.4471 | 1.48234 | 138 | | 0.9496 | 0.9492 | 1.41324 | 21 |
| 0.4961 | 0.4942 | 1.47597 | 125 | | | | | |

Table 7

Volume fraction of alkan-1-ol ($\phi_1$) and temperature derivative of the excess relative permittivity at frequency $f = 1$ MHz (($\partial \varepsilon_r^E / \partial T)_p$) of alkan-1-ol (1) + benzylamine (2) liquid mixtures as functions of the mole fraction of the alkan-1-ol ($x_1$) at temperature $T$ and pressure $p = 0.1$ MPa. [a]

| $x_1$ | $\phi_1$ | $(\partial \varepsilon_r^E / \partial T)_p$ | $x_1$ | $\phi_1$ | $(\partial \varepsilon_r^E / \partial T)_p$ |
|---|---|---|---|---|---|
| \multicolumn methanol (1) + benzylamine (2) ; $T$/K = 298.15 | | | | | |
| 0.0491 | 0.0188 | 0.0003 | 0.5528 | 0.3148 | -0.0056 |
| 0.1123 | 0.0449 | 0.0004 | 0.5952 | 0.3534 | -0.0075 |
| 0.1520 | 0.0625 | 0.0004 | 0.6515 | 0.4100 | -0.0092 |
| 0.2044 | 0.0872 | 0.0002 | 0.6944 | 0.4579 | -0.0102 |
| 0.2499 | 0.1102 | 0.0002 | 0.7498 | 0.5270 | -0.0111 |
| 0.2989 | 0.1368 | -0.0001 | 0.7983 | 0.5953 | -0.0106 |
| 0.3617 | 0.1740 | -0.0013 | 0.8516 | 0.6808 | -0.0094 |
| 0.3922 | 0.1935 | -0.0018 | 0.8991 | 0.7681 | -0.0086 |
| 0.4506 | 0.2336 | -0.0031 | 0.9500 | 0.8760 | -0.0021 |
| 0.4988 | 0.2700 | -0.0044 | | | |
| propan-1-ol (1) + benzylamine (2) ; $T$/K = 298.15 | | | | | |
| 0.0515 | 0.0359 | 0.0019 | 0.5509 | 0.4570 | 0.0091 |
| 0.1016 | 0.0720 | 0.0033 | 0.6023 | 0.5096 | 0.0075 |
| 0.1494 | 0.1076 | 0.0049 | 0.6509 | 0.5613 | 0.0069 |
| 0.2006 | 0.1469 | 0.0056 | 0.7017 | 0.6175 | 0.0044 |
| 0.2448 | 0.1820 | 0.0077 | 0.7489 | 0.6718 | 0.0037 |
| 0.3011 | 0.2282 | 0.0080 | 0.7999 | 0.7328 | 0.0013 |
| 0.3532 | 0.2726 | 0.0088 | 0.8496 | 0.7949 | 0.0003 |
| 0.4003 | 0.3141 | 0.0090 | 0.9000 | 0.8606 | -0.0012 |
| 0.4505 | 0.3600 | 0.0097 | 0.9497 | 0.9283 | -0.0005 |
| 0.5007 | 0.4076 | 0.0088 | | | |
| pentan-1-ol (1) + benzylamine (2) ; $T$/K = 298.15 | | | | | |
| 0.0513 | 0.0509 | 0.0031 | 0.5485 | 0.5466 | 0.0240 |
| 0.0974 | 0.0967 | 0.0056 | 0.5991 | 0.5973 | 0.0234 |
| 0.1492 | 0.1482 | 0.0088 | 0.6480 | 0.6463 | 0.0234 |
| 0.1986 | 0.1974 | 0.0111 | 0.6982 | 0.6966 | 0.0213 |
| 0.2499 | 0.2485 | 0.0138 | 0.7498 | 0.7484 | 0.0196 |
| 0.2978 | 0.2962 | 0.0161 | 0.7986 | 0.7974 | 0.0154 |
| 0.3478 | 0.3461 | 0.0186 | 0.8484 | 0.8474 | 0.0128 |
| 0.3999 | 0.3981 | 0.0203 | 0.8991 | 0.8984 | 0.0071 |
| 0.4490 | 0.4471 | 0.0222 | 0.9496 | 0.9492 | 0.0041 |
| 0.4961 | 0.4942 | 0.0226 | | | |

[a] Standard uncertainties, $u$: $u(T) = 0.02$ K; $u(p) = 10$ kPa; $u(f) = 20$ Hz. Expanded uncertainties, $U$ (coverage factor = 2, approx. 95% confidence level): $U(x_1) = 0.0010$; $U(\phi_1) = 0.004$, $U\left[(\partial \varepsilon_r^E / \partial T)_p\right] = 0.0008$ K$^{-1}$.

Table 8

Coefficients $A_i$ and standard deviations, $\sigma(F^E)$ (equation (12)), for the representation of $F^E$ at temperature $T$ and pressure $p = 0.1$ MPa for alkan-1-ol (1) + benzylamine (2) liquid mixtures by equation (11).

| Property $F^E$ | alkan-1-ol | $T/K$ | $A_0$ | $A_1$ | $A_2$ | $A_3$ | $A_4$ | $A_5$ | $A_6$ | $\sigma(F^E)$ |
|---|---|---|---|---|---|---|---|---|---|---|
| $V_m^E$ /cm$^3\cdot$mol$^{-1}$ | methanol | 293.15 | -6.095 | -1.97 | -0.26 | | | | | 0.004 |
| | | 298.15 | -6.17 | -1.87 | -0.36 | | | | | 0.007 |
| | | 303.15 | -6.202 | -2.00 | -0.38 | | | | | 0.003 |
| | propan-1-ol | 293.15 | -4.190 | -0.915 | 0.09 | | | | | 0.0018 |
| | | 298.15 | -4.138 | -0.91 | 0.11 | | | | | 0.004 |
| | | 303.15 | -4.156 | -0.85 | 0.09 | | | | | 0.003 |
| | pentan-1-ol | 293.15 | -3.096 | -0.39 | 0.20 | 0.27 | | | | 0.002 |
| | | 298.15 | -3.06 | -0.31 | 0.26 | | | | | 0.006 |
| | | 303.15 | -3.026 | -0.35 | 0.35 | 0.43 | | | | 0.004 |
| $\kappa_S^E$ /TPa$^{-1}$ | methanol | 298.15 | -439.1 | -461 | -362 | -203 | -80 | -162 | -148 | 0.15 |
| | propan-1-ol | 298.15 | -284.0 | -163.2 | -71 | -24 | | | | 0.13 |
| | pentan-1-ol | 298.15 | -170.2 | -30.7 | | | | | | 0.11 |
| $c^E$ /m$\cdot$s$^{-1}$ | methanol | 298.15 | 523.9 | 281.8 | 74 | | | | | 0.3 |
| | propan-1-ol | 298.15 | 320.3 | 28.2 | | | | | | 0.17 |
| | pentan-1-ol | 298.15 | 190.4 | -44.6 | 6.5 | | | | | 0.11 |
| $\varepsilon_r^E$ | methanol | 293.15 | -0.17 | 5.3 | 5.2 | 1.2 | | | | 0.02 |
| | | 298.15 | -0.27 | 5.0 | 5.0 | 1.2 | | | | 0.02 |
| | | 303.15 | -0.35 | 4.6 | 4.8 | 1.3 | | | | 0.02 |
| | propan-1-ol | 293.15 | -6.08 | -1.89 | 0.87 | 1.19 | | | | 0.006 |
| | | 298.15 | -5.90 | -1.98 | 0.68 | 1.1 | | | | 0.007 |
| | | 303.15 | -5.71 | -2.04 | 0.51 | 1.0 | | | | 0.008 |
| | pentan-1-ol | 293.15 | -7.864 | -4.15 | -1.1 | 1.06 | 1.0 | | | 0.005 |
| | | 298.15 | -7.408 | -3.97 | -1.15 | 0.87 | 0.9 | | | 0.003 |
| | | 303.15 | -6.939 | -3.78 | -1.20 | 0.7 | 0.8 | | | 0.003 |
| $10^5 n_D^E$ | methanol | 293.15 | 3038 | 1183 | | | | | | 24 |
| | | 298.15 | 2783 | 1746 | | | | | | 22 |
| | | 303.15 | 2812 | 1226 | | | | | | 20 |
| | propan-1-ol | 293.15 | 1323 | 321 | | | | | | 12 |
| | | 298.15 | 1270 | 113 | | | | | | 8 |
| | | 303.15 | 1259 | 156 | | | | | | 11 |
| | pentan-1-ol | 293.15 | 599 | -180 | | | | | | 9 |
| | | 298.15 | 544 | -88 | | | | | | 7 |
| | | 303.15 | 517 | | | | | | | 7 |
| $(\partial\varepsilon_r^E/\partial T)_p$ /K$^{-1}$ | methanol | 298.15 | -0.019 | -0.057 | -0.036 | | | | | 0.0007 |
| | propan-1-ol | 298.15 | 0.0368 | -0.022 | -0.037 | | | | | 0.0005 |
| | pentan-1-ol | 298.15 | 0.0931 | 0.037 | -0.024 | -0.036 | | | | 0.0004 |

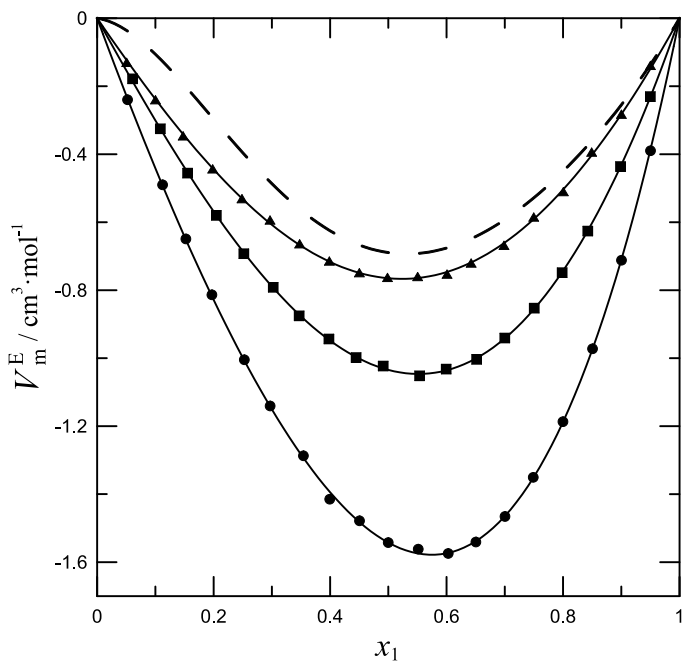

Figure 1. Excess molar volume ($V_m^E$) of alkan-1-ol (1) + benzylamine (2) liquid mixtures as a function of the alkan-1-ol mole fraction ($x_1$) at 298.15 K and 0.1 MPa. Symbols, experimental values (this work): (●), methanol; (■), propan-1-ol; (▲), pentan-1-ol. Solid lines, calculations with equation (11) using the coefficients from Table 8. Dashed line, pentan-1-ol (smoothed curve from reference [67]).

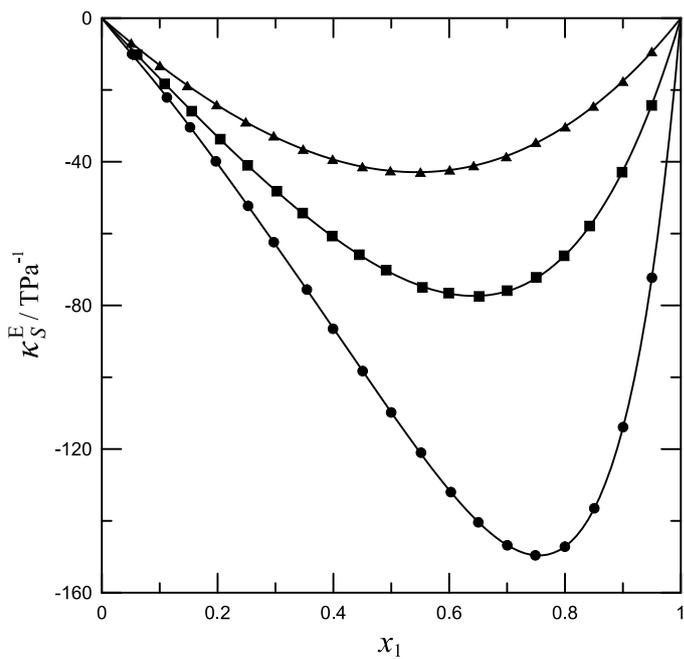

Figure 2. Excess isentropic compressibility ($\kappa_S^E$) of alkan-1-ol (1) + benzylamine (2) liquid mixtures as a function of the alkan-1-ol mole fraction ($x_1$) at 298.15 K and 0.1 MPa. Full symbols, experimental values (this work): (●), methanol; (■), propan-1-ol; (▲), pentan-1-ol. Solid lines, calculations with equation (11) using the coefficients from Table 8.

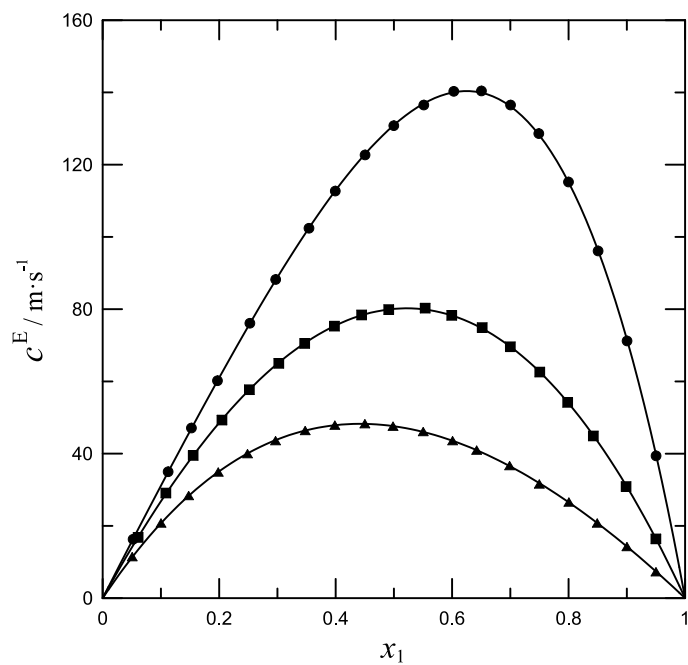

Figure 3. Excess speed of sound ($c^E$) of alkan-1-ol (1) + benzylamine (2) liquid mixtures as a function of the alkan-1-ol mole fraction ($x_1$) at 298.15 K and 0.1 MPa. Full symbols, experimental values (this work): (●), methanol; (■), propan-1-ol; (▲), pentan-1-ol. Solid lines, calculations with equation (11) using the coefficients from Table 8.

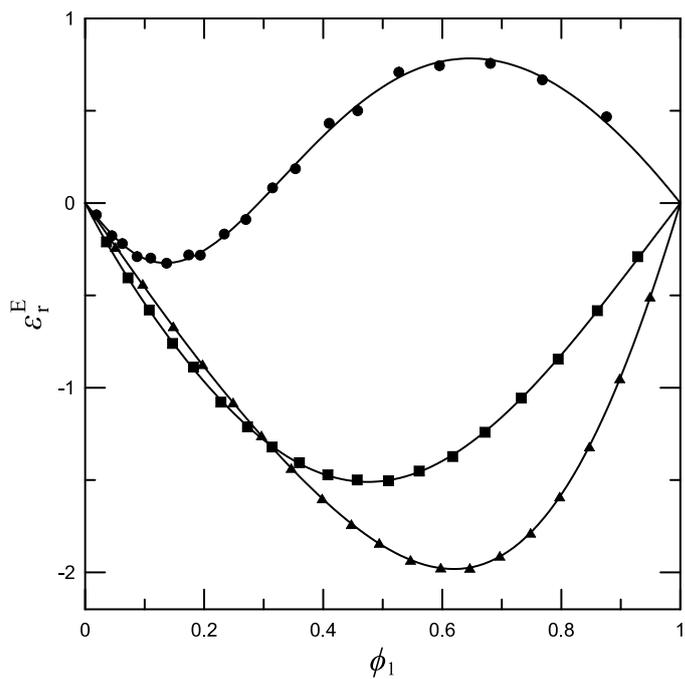

Figure 4. Excess relative permittivity ($\varepsilon_r^E$) of alkan-1-ol (1) + benzylamine (2) liquid mixtures as a function of the alkan-1-ol volume fraction ($\phi_1$) at 298.15 K, 0.1 MPa, and 1 MHz. Symbols, experimental values (this work): (●), methanol; (■), propan-1-ol; (▲), pentan-1-ol. Solid lines, calculations with equation (11) using the coefficients from Table 8.

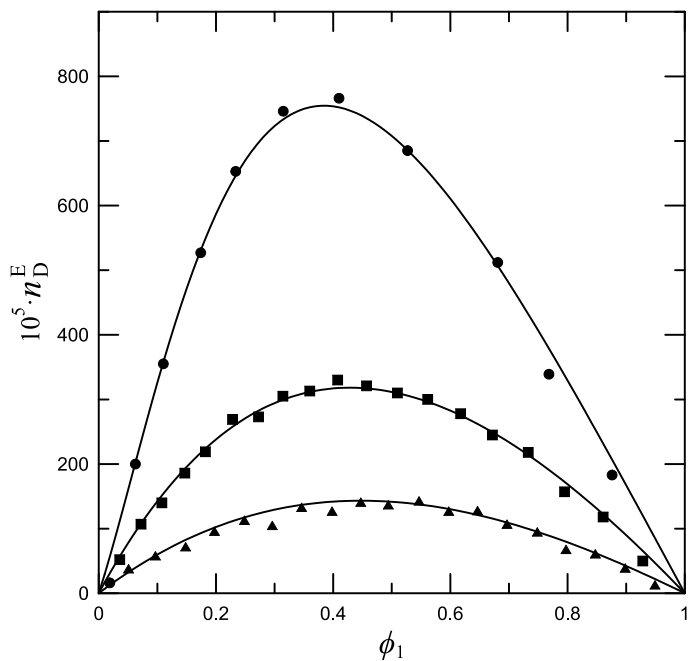

Figure 5. Excess refractive index at the sodium D-line ($n_D^E$) of alkan-1-ol (1) + benzylamine (2) liquid mixtures as a function of the alkan-1-ol volume fraction ($\phi_1$) at 298.15 K and 0.1 MPa. Full symbols, experimental values (this work): (●), methanol; (■), propan-1-ol; (▲), pentan-1-ol. Solid lines, calculations with equation (11) using the coefficients from Table 8.

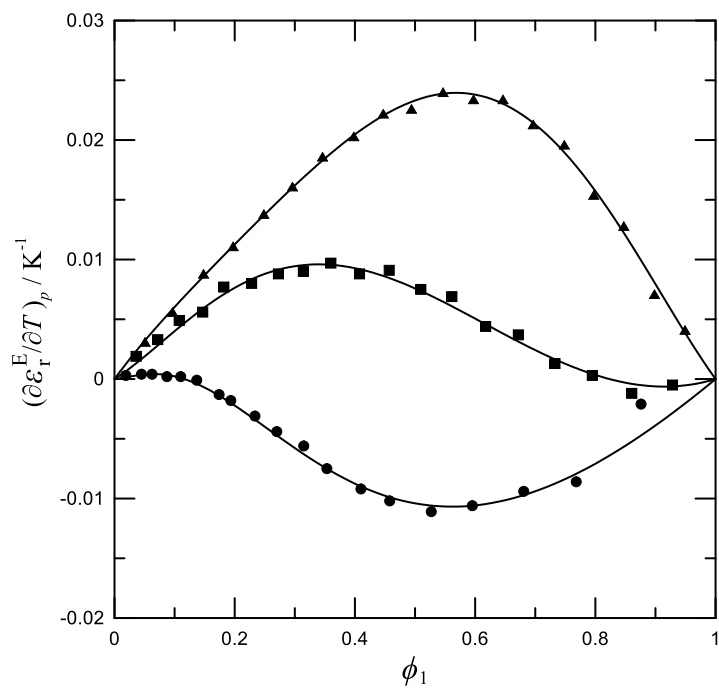

Figure 6. Temperature derivative of the excess relative permittivity ($(\partial \varepsilon_r^E / \partial T)_p$) of alkan-1-ol (1) + benzylamine (2) liquid mixtures as a function of the alkan-1-ol volume fraction ($\phi_1$) at 298.15 K, 0.1 MPa, and 1 MHz. Full symbols, experimental values (this work): ($\bullet$), methanol; ($\blacksquare$), propan-1-ol; ($\blacktriangle$), pentan-1-ol. Solid lines, calculations with equation (11) using the coefficients from Table 8.

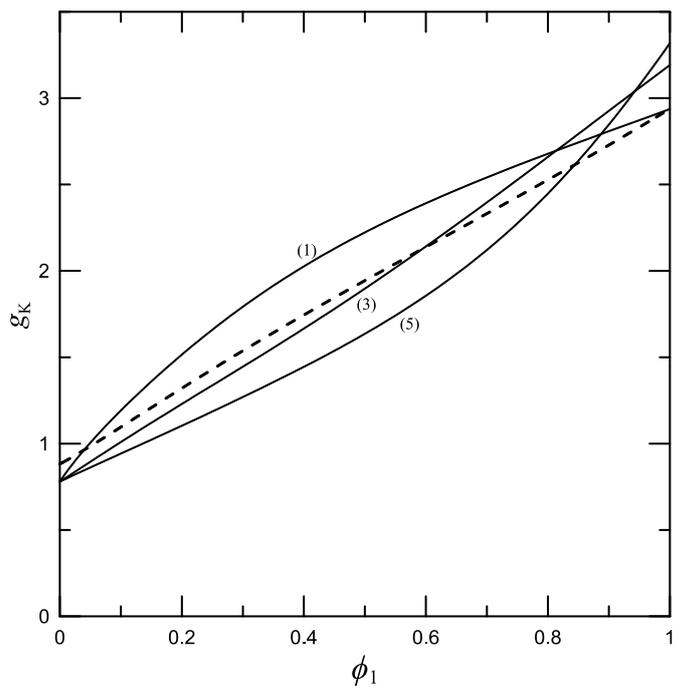

Figure 7. Kirkwood correlation factor ($g_K$) of alkan-1-ol (1) + amine (2) liquid mixtures as a function of the volume fraction of compound (1), $\phi_1$, at 0.1 MPa and 298.15 K: Solid lines, alkan-1-ol (1) + benzylamine (2) (this work); the numbers in parentheses indicate the number of carbon atoms of the alkan-1-ol. (– – –), methanol (1) + aniline (2) ([15]).